\documentclass[aps,prb,twocolumn,superscriptaddress,notitlepage]{revtex4-1}
\usepackage{graphicx}
\usepackage[caption=false]{subfig}
\usepackage{float}
\usepackage{natbib}
\usepackage{xcolor}
\usepackage{xr}
\usepackage{amsmath}
\usepackage{amssymb}
\usepackage{array}
\usepackage{wasysym}
\usepackage{color,soul}
\usepackage{braket}
\usepackage{verbatim}
\usepackage{hyperref}
\usepackage{gensymb}
\usepackage{url}
\usepackage{dirtytalk}
\usepackage{upgreek}

\usepackage{filecontents}
\begin{document}

\title{Discovery of High-Temperature Charge Order and Time-Reversal Symmetry-Breaking in the Kagome Superconductor YRu$_{3}$Si$_{2}$}

\author{P. Král}
\thanks{These authors contributed equally to the experiments.}
\affiliation{PSI Center for Neutron and Muon Sciences CNM, 5232 Villigen PSI, Switzerland}

\author{J.N. Graham}
\thanks{These authors contributed equally to the experiments.}
\affiliation{PSI Center for Neutron and Muon Sciences CNM, 5232 Villigen PSI, Switzerland}

\author{V. Sazgari}
\thanks{These authors contributed equally to the experiments.}
\affiliation{PSI Center for Neutron and Muon Sciences CNM, 5232 Villigen PSI, Switzerland}

\author{I. Plokhikh}
\affiliation{PSI Center for Neutron and Muon Sciences CNM, 5232 Villigen PSI, Switzerland}

\author{A.~Lukovkina}
\affiliation{Department of Quantum Matter Physics, University of Geneva, CH-1211 Geneva, Switzerland}

\author{O. Gerguri}
\affiliation{PSI Center for Neutron and Muon Sciences CNM, 5232 Villigen PSI, Switzerland}

\author{I.~Bia\l{}o}
\affiliation{Physik-Institut, Universit\"{a}t Z\"{u}rich, Winterthurerstrasse 190, CH-8057 Z\"{u}rich, Switzerland}
\affiliation{AGH University of Science and Technology, Faculty of Physics and Applied Computer Science, 30-059 Krak\'{o}w, Poland}

\author{A. Doll}
\affiliation{PSI Center for Neutron and Muon Sciences CNM, 5232 Villigen PSI, Switzerland}

\author{L. Martinelli}
\affiliation{Physik-Institut, Universit\"{a}t Z\"{u}rich, Winterthurerstrasse 190, CH-8057 Z\"{u}rich, Switzerland}

\author{J. Oppliger}
\affiliation{Physik-Institut, Universit\"{a}t Z\"{u}rich, Winterthurerstrasse 190, CH-8057 Z\"{u}rich, Switzerland}

\author{S.S. Islam}
\affiliation{PSI Center for Neutron and Muon Sciences CNM, 5232 Villigen PSI, Switzerland}

\author{K. Wang}
\affiliation{Department of Materials Science and Metallurgy, University of Cambridge, Cambridge, UK}

\author{M.~Salamin}
\affiliation{Department of Quantum Matter Physics, University of Geneva, CH-1211 Geneva, Switzerland}

\author{H. Luetkens}
\affiliation{PSI Center for Neutron and Muon Sciences CNM, 5232 Villigen PSI, Switzerland}

\author{R. Khasanov}
\affiliation{PSI Center for Neutron and Muon Sciences CNM, 5232 Villigen PSI, Switzerland}

\author{M.v. Zimmermann}
\affiliation{Deutsches Elektronen-Synchrotron, 22607 Hamburg, Germany}

\author{J.-X.~Yin}
\affiliation{Department of Physics, Southern University of Science and Technology, Shenzhen, Guangdong, 518055, China}

\author{Ziqiang Wang}
\affiliation{Department of Physics, Boston College, Chestnut Hill, Massachusetts 02467, USA}

\author{J.~Chang}
\affiliation{Physik-Institut, Universit\"{a}t Z\"{u}rich, Winterthurerstrasse 190, CH-8057 Z\"{u}rich, Switzerland}

\author{B. Monserrat}
\affiliation{Department of Materials Science and Metallurgy, University of Cambridge, Cambridge, UK}

\author{D. Gawryluk}
\affiliation{PSI Center for Neutron and Muon Sciences CNM, 5232 Villigen PSI, Switzerland}

\author{F.O.~von~Rohr}
\affiliation{Department of Quantum Matter Physics, University of Geneva, CH-1211 Geneva, Switzerland}

\author{S.-W. Kim}
\email{swk38@cam.ac.uk}
\affiliation{Department of Materials Science and Metallurgy, University of Cambridge, Cambridge, UK}

\author{Z. Guguchia}
\email{zurab.guguchia@psi.ch}
\affiliation{PSI Center for Neutron and Muon Sciences CNM, 5232 Villigen PSI, Switzerland}

\date{\today}


\maketitle

\textbf{The identification of high-temperature unconventional charge order and superconductivity in kagome quantum materials is pivotal for deepening our understanding of geometrically frustrated and correlated electron systems, and for harnessing their exotic properties in future quantum technologies. Here, we report the discovery of a remarkably rich phase diagram in the kagome superconductor YRu$_{3}$Si$_{2}$, uncovered through a unique combination of muon spin rotation (${\mu}$SR), magnetotransport, X-ray diffraction (XRD), and density functional theory (DFT) calculations. Our study reveals the emergence of a charge-ordered state with a propagation vector of (1/2, 0, 0), setting a record onset temperature of 800 K for such an order in a kagome system and for quantum materials more broadly. In addition, we observe time-reversal symmetry (TRS) breaking below $T_{2}^{*}$ ${\simeq}$ 25 K and field-induced magnetism below $T_{1}^{*}$ ${\simeq}$ 90 K, indicating the presence of a hidden magnetic state. These transitions are mirrored in the magnetoresistance data, which show a clear onset at  ${\sim}$ $T_{1}^{*}$ and a pronounced increase below  ${\sim}$ $T_{2}^{*}$, ultimately reaching a maximum magnetoresistance of 45${\%}$. Band structure calculations identify two van Hove singularities (VHSs) near the Fermi level, one of which resides within a flat band, suggesting a strong interplay between electronic correlations and emergent orders. At low temperatures, we find bulk superconductivity below $T_{\rm c}$ = 3.4 K, characterized by a pairing symmetry with either two isotropic full gaps or an anisotropic nodeless gap. Together, our findings point to a coexistence of high-temperature charge order, tunable magnetism, and multigap superconductivity in YRu$_{3}$Si$_{2}$, positioning it as a compelling platform for exploring strongly correlated kagome physics.}

Identifying novel quantum phases and emergent electronic transitions—especially those intertwined with superconductivity at elevated temperatures—remains one of the central pursuits in condensed matter physics \cite{keimer2015quantum}. The interplay between symmetry breaking, electronic correlations, and topology often gives rise to unconventional states of matter with potential relevance for quantum technologies. Among the diverse platforms studied, kagome lattice systems have recently garnered significant attention as fertile ground for realizing such exotic phenomena.

The kagome lattice \cite{Syozi1951}, with its geometrically frustrated network of corner-sharing triangles, hosts a unique band structure characterized by Dirac crossings, flat bands, and van Hove singularities (VHSs), making it an ideal setting to explore the interplay of electronic topology, magnetism, and superconductivity. A growing family of kagome metals—including AV$_{3}$Sb$_{5}$ (A = K, Rb, Cs), CsCr$_{3}$Sb$_{5}$, CsTi$_{3}$Bi$_{5}$, ScV$_{6}$Sn$_{6}$, FeGe, and Ta$_{2}$V$_{3.1}$Si$_{0.9}$—has exhibited a wealth of correlated behaviors \cite{wilson2024v3sb5,ortiz2019new,ortiz2020cs,PhysRevLett.127.217601,yin2022topological,guguchia2023unconventional,mielke2022time,plokhikh2024discovery,neupert2022charge,YinNM,Wagner2023,PhysRevLett.110.126405,teng2022discovery,liu2024superconductivity,yang2024superconductivity,Christensen2022,Guo2022,BonfaALC,wang2024origin,PhysRevB.107.045127,Graham2024,xu2022three,ingham2025vestigial}, such as charge order (CO), nematicity, time-reversal symmetry (TRS) breaking \cite{guguchia2023unconventional,Guo2022,scammell2023chiral}, and unconventional superconductivity. These findings underscore the potential of the Kagome lattice as a platform that hosts emergent quantum phenomena.

\begin{figure*}
    \centering
    \includegraphics[width=1.0\linewidth]{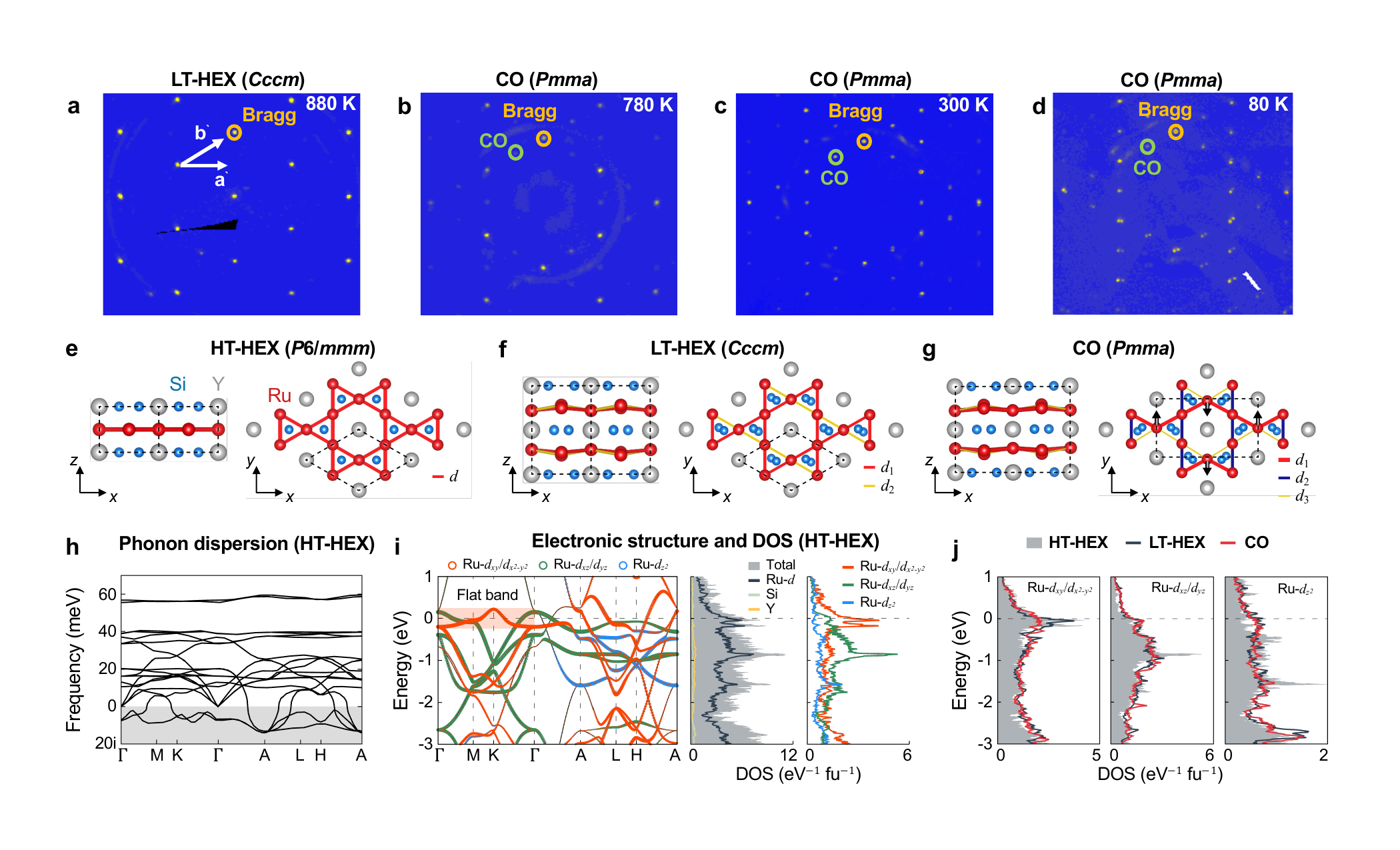}
    \vspace{-1.5cm}
    \caption{\textbf{Room temperature charge order, electronic structure and density of states in YRu$_{3}$Si$_{2}$.} \textbf{a-d}, Reconstructed reciprocal space along the (0 0 1) direction at 2 r.l.u. (reciprocal lattice units), performed at various temperatures $T$ = 80 K, 300 K, 780 K, and 880 K, respectively. Arrows indicate the reciprocal space vectors. Orange and green circles mark the Bragg peak and charge order (CO) peaks, respectively. \textbf{e-g}, Atomic structures of \textbf{e} HT-HEX ($P6/mmm$), \textbf{f} LT-HEX ($Cccm$), and \textbf{g} CO ($Pmma$) phases in YRu$_{3}$Si$_{2}$. Dashed lines indicate the unit cells. In each structure, Ru-Ru bonds of different lengths are depicted in different colors. In \textbf{g}, black arrows in the right panel show the in-plane CO displacement patterns of Ru atoms. \textbf{h}, Phonon dispersion of the HT-HEX structure. \textbf{i}, Orbital-projected band structure and density of states (DOS) of the HT-HEX structure. The $d$ orbitals of Ru atoms are projected onto the band structure, with the radius of the open circles proportional to the projected weight. \textbf{j}, Orbital-projected DOS of the three structures. }
    \label{Figure1}
\end{figure*}

Within this landscape, LaRu$_{3}$Si$_{2}$ \cite{Vandenberg1980,mielke2021nodeless,plokhikh2024discovery,mielke2024chargeordersdistinctmagnetic,ma2024domeshapedsuperconductingphasediagram,misawa2025chemicalenhancementsuperconductivitylaru3si2,deng2025theory} has recently attracted attention due to its rich physics. It exhibits charge order with a record onset temperature of 400 K \cite{plokhikh2024discovery}, followed by a secondary charge order with field-induced magnetism below 80 K and TRS breaking below 35 K \cite{mielke2024chargeordersdistinctmagnetic}. High-pressure studies on LaRu$_{3}$Si$_{2}$ by our group \cite{ma2024domeshapedsuperconductingphasediagram} and others \cite{PhysRevB.111.144505}  reveal a dome-shaped superconducting phase and a similar trend in the normal-state electronic response, suggesting unconventional pairing and a positive correlation between superconductivity and normal-state electronic behavior. These features distinguish it from other kagome materials and motivate the exploration of related compounds within the same structural family.

To further understand the kagome 132 systems, one promising approach is to apply chemical pressure—for example, by substituting La with the smaller Y ion, resulting in YRu$_{3}$Si$_{2}$. While YRu$_{3}$Si$_{2}$ retains the same crystal structure, its superconducting transition temperature ($T_{\rm c}$ ${\simeq}$ 3.4 K) \cite{Gong_2022} is roughly half that of LaRu$_{3}$Si$_{2}$. This raises fundamental questions about the origin of superconductivity and its interplay with other ordering phenomena in this material class. It is therefore of great interest to explore the microscopic nature of superconductivity in YRu$_{3}$Si$_{2}$, particularly focusing on the superfluid density and the symmetry of the superconducting gap. Even more crucially, we seek to determine whether charge order—a prominent feature of LaRu$_{3}$Si$_{2}$—also emerges in YRu$_{3}$Si$_{2}$, and how its presence or absence may be linked to the observed suppression of $T_{\rm c}$. Gaining such insights is vital for deepening our understanding of intertwined electronic orders in kagome superconductors and for guiding the discovery and design of new materials with tunable quantum phases.

In this article, we report the discovery of charge order in YRu$_{3}$Si$_{2}$ with a propagation vector of (1/2, 0, 0) and a record onset temperature of 800 K, alongside field-induced magnetism below 90 K and time-reversal symmetry (TRS) breaking below 25 K. DFT calculations reveal a characteristic kagome band structure featuring two van Hove singularities near the Fermi level, one of which lies within a flat band. Bulk superconductivity emerges below 3.4 K, with a two-gap (s+s)-wave or anisotropic s-wave pairing symmetry. These findings demonstrate a rare coexistence of charge order, magnetism, and multigap superconductivity, establishing YRu$_{3}$Si$_{2}$ as a model system for correlated kagome physics.

\begin{figure*}
    \centering
    \includegraphics[width=1\linewidth]{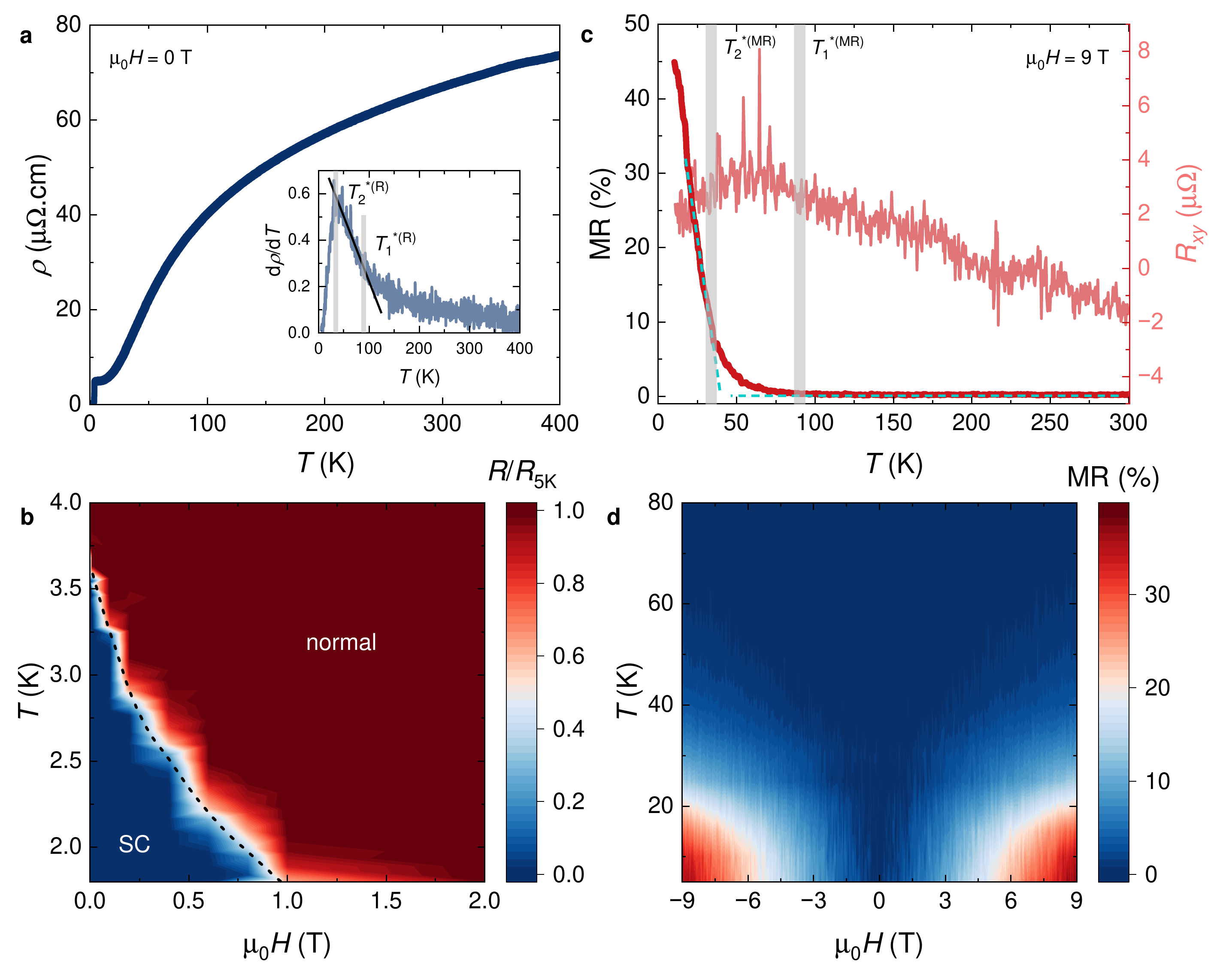}
    \caption{\textbf{Magnetotransport characteristics for YRu$_{3}$Si$_{2}$.} \textbf{a}, Temperature dependence of electrical resistivity and its first derivative (inset). \textbf{b}, Contour map of the resistivity data across the superconducting transition as a function of applied magnetic field. \textbf{c}, Temperature-dependent magnetoresistance (left axis), and the Hall resistance (right axis) obtained at 9 T. Vertical gray lines in \textbf{a} and \textbf{c} mark the characteristic temperatures. \textbf{d}, Temperature-field contour map of magnetoresistance.}
    \label{fig:2}
\end{figure*}

Firstly, to determine the presence of charge order, X-ray diffraction (XRD) experiments on YRu$_{3}$Si$_{2}$ were conducted over a temperature range of 80 K to approximately 900 K. The reconstructed reciprocal patterns obtained at various temperatures are shown in Fig. \ref{Figure1}a-d. Above 800 K, we identified the $Cccm$ structure, the same phase as observed in LaRu$_{3}$Si$_{2}$ \cite{plokhikh2024discovery} between 400 K and 600 K. Below 800 K, down to 80 K, superstructure reflections emerge, which in the hexagonal setting correspond to \textbf{q}$_{1}$ = (1/2, 0, 0), \textbf{q}$_{2}$ = (0, 1/2, 0), and \textbf{q}$_{3}$ = (1/2, $-$1/2, 0). In the orthorhombic setting, these reflections indicate a breaking of the $C$-centering, signifying a structural transition from $Cccm$ to $Pmma$. This transition establishes charge order at $T_{\rm co}$ = 800 K in YRu$_{3}$Si$_{2}$, which is twice as high as in LaRu$_{3}$Si$_{2}$ ($T_{\rm co}$ = 400 K), setting a new record for charge order in kagome systems-and in quantum materials more broadly. Note, that unlike for the case of LaRu$_{3}$Si$_{2}$, the $P6/mmm$ (HT-HEX) was not observed experimentally up to the highest temperature used in our measurements. Still, presence of the hexagonal symmetry and the necessity of using the hexagonal twin domains for structure refinement suggests the existence of the high-temperature phase with hexagonal symmetry.\par

To gain a more detailed insight into the distortions and phase stability of YRu$_{3}$Si$_{2}$, first-principles DFT calculations have been performed. Our calculations identify the experimentally observed $Cccm$ and $Pmma$ structures, and we consider their parent high-temperature $P6/mmm$ structure (Fig. \ref{Figure1}e-j). The parent structure $P6/mmm$, with an undistorted kagome lattice of Ru atoms, exhibits multiple unstable imaginary modes across the whole BZ in its phonon dispersion (Fig. \ref{Figure1}h). The imaginary phonon mode at the A point gives rise to the $Cccm$ structure characterized by in-plane displacements of Si atoms and out-of-plane displacements of Ru atoms. The out-of-plane modulation yields two different bond lengths in the Ru kagome net, $d_1=2.759\,\text{\AA}$ and $d_2=2.793\,\text{\AA}$, both of which are larger than $d=2.736\,\text{\AA}$ in the parent $P6/mmm$ structure. This structural distortion lifts peaks in the density of states (DOS) of the out-of-plane $d_{xz}$, $d_{yz}$, and $d_{z^2}$ orbitals far below the Fermi level (Fig. \ref{Figure1}j). The large DOS at the Fermi level, originating from the flat band of the in-plane $d_{xy}$ and $d_{x^2-y^2}$ orbitals (Fig. \ref{Figure1}i), remains nearly unchanged upon the distortion. For the $Pmma$ structure, the structural distortion arises from imaginary phonon modes at A and L points, featuring in-plane Ru distortions (see arrows in Fig. \ref{Figure1}g) in addition to in-plane Si distortions and
out-of-plane Ru distortions. The bond lengths between Ru atoms are $d_1=2.708\,\text{\AA}$, $d_2=2.760\,\text{\AA}$, and $d_3=2.833\,\text{\AA}$, with $d_1$ being shorter than $d$ in the $P6/mmm$ structure, demonstrating its charge order characteristics. The charge order nature of the $Pmma$ structure is further confirmed by the reduction in the large DOS of the in-plane $d_{xy}$ and $d_{x^2-y^2}$ orbitals at the Fermi level, compared to the parent $P6/mmm$ structure (Fig. \ref{Figure1}j).

\begin{figure*}
    \centering
    \includegraphics[width=1\linewidth]{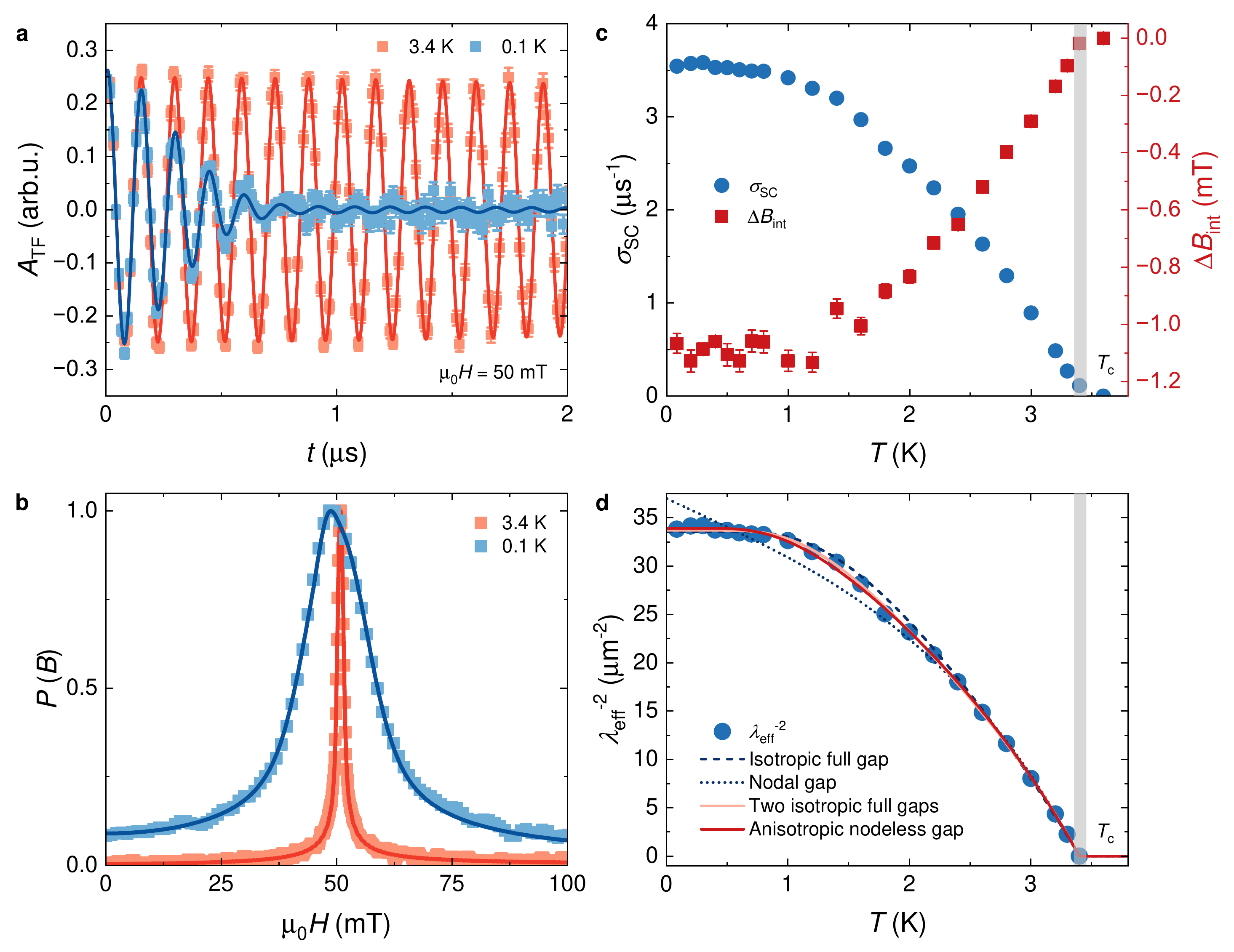}
    \caption{\textbf{Superconducting-state properties of YRu$_{3}$Si$_{2}$ probed by $\mu$SR.} \textbf{a}, Transverse-field (TF) $\mu$SR time spectra obtained above (3.4 K) and below (0.1 K) the SC transition under an applied magnetic field of 50 mT. The solid lines represent fits to the data by means of Eq.~\eqref{eqS1}. \textbf{b}, Normalized Fourier transform of the measured time spectra in the normal and superconducting state, respectively. \textbf{c}, Temperature dependence of the muon spin depolarization rate \(\sigma_\text{sc}(T)\) (left axis), and diamagnetic shift \(\Delta B_\text{int}(T)\) (right axis), at 50 mT. \textbf{d}, Inverse squared effective penetration depth \(\lambda_\text{eff}^{-2}\) as a function of temperature fitted with several theoretical models (see Methods section).}
    \label{fig:3}
\end{figure*}

\begin{figure*}
    \centering
    \includegraphics[width=1\linewidth]{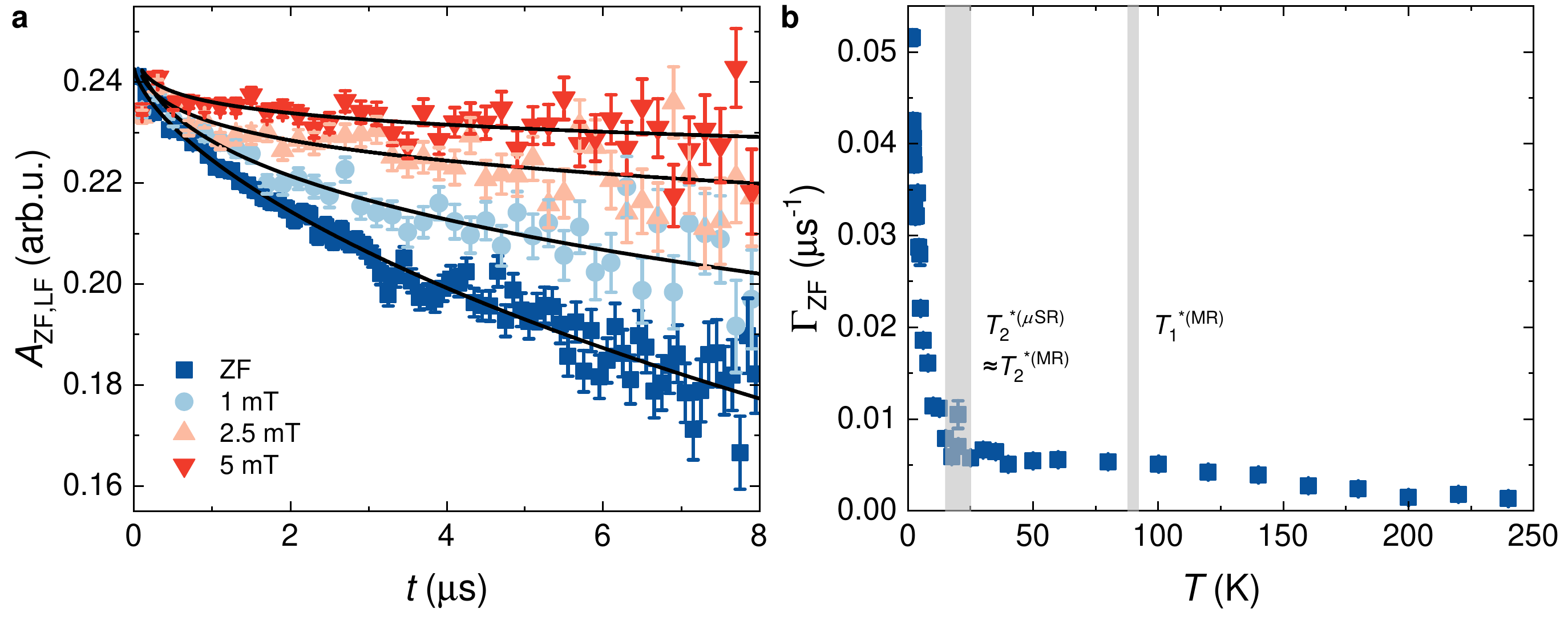}
    \caption{\textbf{Zero-field (ZF) and longitudinal-field (LF) $\mu$SR study on YRu$_{3}$Si$_{2}$.} \textbf{a}, ZF and LF $\mu$SR time spectra recorded at 5 K. The black solid line represents a fit using a simple exponential function. \textbf{b}, Temperature dependence of ZF muon spin relaxation rate \(\Gamma_\text{ZF}(T)\). Vertical grey lines indicate the onset temperature $T_{1}^{*\text{(MR)}}$, below which MR emerges, the temperature $T_{2}^{*\text{(MR)}}$, below which a more pronounced increase in MR is observed and the temperature $T_{2}^{*\text{($\mu$SR)}}$, below which ZF muon spin relaxation rate increases. Notably, $T_{2}^{*\text{(MR)}}$ ${\simeq}$ $T_{2}^{*\text{($\mu$SR)}}$.}
    \label{fig:4}
\end{figure*}

Having observed charge order and confirmed it through DFT calculations, we next investigate the transport characteristics in both the superconducting and normal states.
The temperature dependence of electrical resistivity for YRu$_{3}$Si$_{2}$ depicted in Fig. \ref{fig:2}a establishes three temperature scales governing the electronic behavior. Inspecting the first derivative of the resistivity data reveals a subtle change in slope at \(T_{1}^{*\text{(R)}}\sim90\text{ K}\), followed by a pronounced peak at \(T_{2}^{*\text{(R)}}\sim35\text{ K}\). These values are comparable to \(T_\text{CO,II}\simeq80\text{ K}\) and \(T^*\simeq35\text{ K}\) observed in LaRu$_{3}$Si$_{2}$. In the low-temperature range, the superconducting state is reached with the onset and midpoint (50\% drop in resistivity) at 3.9 K and 3.7 K, respectively. Superconducting transitions under varying magnetic fields are presented in the color plot, with temperature on the vertical axis and magnetic field on the horizontal axis. Blue regions represent the zero-resistance superconducting state, while red corresponds to the normal (resistive) state. The application of a magnetic field results in the gradual suppression of the transition temperature (Fig. \ref{fig:2}b) with an estimated upper critical field \(\mu_0H_\text{c2}=1.0\text{ T}\) at \(T=2\text{ K}\).
Motivated by the presence of two distinct temperature scales observed in the normal state via the resistivity derivative, magnetoresistance (MR) was measured across a broad range of temperatures and magnetic fields. The resulting data, displayed in the color plot (\ref{fig:2}d), reveal that MR emerges below approximately 80 K and reaches up to 45${\%}$. The temperature dependence of the MR measured at 9 T is presented in Fig. \ref{fig:2}c. Two clear slope changes are observed, showing the initial deviation from zero MR below \(T_{1}^{*\text{(MR)}}\) with a significant increase below \(T_{2}^{*\text{(MR)}}\) which coincides with the values determined from the derivative of resistivity. Additionally, the temperature dependence of the Hall resistivity shows a broad anomaly in the corresponding temperature range. Our MR experiments therefore indicate the presence of a transition or crossover in the normal state, occurring between the onset of charge order and the emergence of superconductivity.

In order to unveil the microscopic nature of both normal and superconducting states of YRu$_{3}$Si$_{2}$, ${\mu}$SR experiments were carried out. Unlike many other methods that probe the penetration depth (${\lambda}$) only near the surface, the ${\mu}$SR technique offers a powerful means of measuring the superfluid density in the vortex state of type-II superconductors deep within the bulk of the sample. Further details are provided in the “Methods” section. Additionally, zero-field ${\mu}$SR can detect internal magnetic fields as small as 0.1 G without the need for an external field, making it an especially valuable tool for probing spontaneous magnetism associated with TRS breaking. Fig. \ref{fig:3}a illustrates the transverse-field (TF) ${\mu}$SR time spectra in an applied magnetic field of 50 mT above (3.4 K) and below (0.1 K) the SC transition. Above \(T_\text{c}\), the oscillations show a weak damping due to the random local fields from the nuclear moments, while below \(T_\text{c}\) the damping rate strongly increases indicating the presence of a non-uniform local magnetic field distribution due to the formation of a flux-line lattice. Further evidence is provided in Fig. \ref{fig:3}b showing the Fourier transform of the ${\mu}$SR spectra at corresponding temperatures. Above $T_{\rm c}$, a sharp and symmetric peak is observed. In the superconducting state, however, the field distribution becomes significantly broadened and asymmetric—features characteristic of a vortex lattice—and is shifted away from the applied magnetic field. The diamagnetic shift \(\Delta B_\text{int}=\mu_0(H_\text{int, SC}-H_\text{int, NS})\), i.e., the difference between the applied field and the central field in a superconducting state, as a function of temperature is shown in Fig. \ref{fig:3}c. The large diamagnetic response of 1.1 mT that is observed is associated with the superconducting transition at \(T_\text{c}=3.4\text{ K}\). This large diamagnetic shift indicates the bulk character of superconductivity. An in-depth look at the superconducting properties is offered by the analysis of the muon spin depolarization rate \(\sigma_\text{tot}\) (\(=\sqrt{\sigma_\text{SC}^2+\sigma_\text{nm}^2}\)), consisting of superconducting, \(\sigma_\text{SC}\), and nuclear magnetic dipolar, \(\sigma_\text{nm}\), contributions. To estimate the superconducting relaxation rate \(\sigma_\text{SC}\), nuclear contribution was considered to be constant above \(T_\text{c}\) and subtracted accordingly. The temperature dependence of $\sigma
_{\rm SC}$ of YRu$_{3}$Si$_{2}$ is shown in \ref{fig:3}c, which was extracted using the equations described in the Method section. The form of the temperature dependence of $\sigma_{\rm SC}$, which reflects the topology of the SC gap, shows saturation below $T$/$T_{\rm c}$ ${\simeq}$ 0.3. We show in the following how these behaviors indicate a nodeless SC gap.\par

For a perfect triangular lattice, the relaxation rate is directly linked to the magnetic penetration depth \(\lambda_\text{eff}\) according to the equation\cite{brandt1988flux}:
\[\frac{\sigma_\text{SC}(T)}{\gamma_{\mu}}=0.06091\frac{\Phi_{0}}{\lambda_\text{eff}^2(T)},\]
where \(\gamma_{\mu}\) is the gyromagnetic ratio of the muon and \(\Phi_{0}\) stands for the magnetic-flux quantum. The magnetic penetration depth is one of the most fundamental parameters in a superconductor since it is related to the superfluid density, $n_s$, via 1/${\lambda}^{2}$=${\mu}_{0}$$e^{2}$$n_{\rm s}$/$m^{*}$ (where $m^*$ is the effective mass). The temperature dependence of 1/${\lambda}^{2}$ is shown in Fig. \ref{fig:3}d. To enable a quantitative analysis, the experimental data were fitted using theoretical models corresponding to a single isotropic full gap, two isotropic full gaps, an anisotropic nodeless gap, and a nodal gap. The fits are presented in Fig. \ref{fig:3}d. The nodal gap model is clearly incompatible with the experimental data, and the single isotropic full-gap model also fails to adequately describe the temperature dependence of \(\lambda_\text{eff}^{-2}(T)\). In contrast, both the two-gap isotropic model and the anisotropic nodeless gap model yield fits of comparable and satisfactory quality. The first approach suggests the presence of two SC gaps, \(\Delta_1=0.53(1)\text{ meV}\) and \(\Delta_2=0.15(1)\text{ meV}\) with a relative weights of 0.93(1) and 0.07(1). In the second scenario, the superconducting gap exhibits an angular dependence similarly as in the case of nodal d-wave superconductivity, however, without reaching the zero value at any point. The anisotropy ratio of the minimum gap value to the maximum \(\Delta_1=0.52(1)\text{ meV}\) is \(a=0.20(2)\), which is lower compared to other anisotropic s-wave kagome superconductors, e.g., CeRu$_{2}$\cite{mielke2022local}. The SC gap structure for the La analog, LaRu$_{3}$Si$_{2}$, was described by a single s-wave model\cite{mielke2021nodeless}, nevertheless, the presence of two gaps cannot be excluded, as well\cite{ushioda2024two}, with a similar weight ratio (0.1 and 0.9) as observed here. The magnetic penetration depth extrapolated to zero temperature is estimated to be $\lambda_{eff}(0)$ = 173(3) nm. The ratio between the superconducting critical temperature and the zero-temperature superfluid density, $T_\text{c}/\lambda_{eff}(0)^{-2}$, is approximately 0.10 for YRu$_{3}$Si$_{2}$, which is lower than the value of ~0.37 reported for LaRu$_{3}$Si$_{2}$\cite{mielke2021nodeless}, but still within the typical range for unconventional superconductors (0.1–20) \cite{uemura1989universal}. In contrast, conventional BCS superconductors exhibit significantly smaller values. This indicates that the superfluid density in YRu$_{3}$Si$_{2}$ is relatively dilute and points toward an unconventional pairing mechanism.\par




Having established the superconducting gap structure and superfluid density in YRu$_{3}$Si$_{2}$, we now shift our focus to the normal state. A key question is whether magnetism is associated with the temperature scales $T_{1}^{*}$ and $T_{2}^{*}$, identified from our transport measurements. In the following, we present results from the powerful combination of zero-field and high transverse-field ${\mu}$SR experiments to address this question. The zero-field (ZF)-$\mu$SR spectrum (Fig. \ref{fig:4}a) is characterized by a weak depolarization of the muon spin ensemble, indicating no evidence of long-range ordered magnetism in YRu$_{3}$Si$_{2}$. However, the muon spin relaxation has a clearly observable temperature dependence. Since the full polarization can be recovered  by the application of a small external longitudinal magnetic field, $B_{{\rm LF}}$~=~5~mT, the relaxation is, therefore, due to spontaneous fields which are static on the microsecond timescale. The zero-field ${\mu}$SR spectra for YRu$_{3}$Si$_{2}$ were fitted using the simple exponential function $P_\text{ZF}(t)$ = exp(-$\Gamma_\text{ZF}t$). Across $T_{1}^{*}$, there is only a change in the slope of ${\Gamma}$. However, a significant observation occurs as the temperature is lowered below $T_{2}^{*}$ ${\simeq}$ 25 K, where there is a notable increase in $\Gamma_\text{ZF}$ (Fig. \ref{fig:4}b. Keeping in mind that there is no structural distortion across $T_{2}^{*}$, we can dismiss changes in the structure being the origin for the increase of relaxation rate. Therefore, we interpret our ZF-$\mu$SR results as an indication that there is an enhanced width of internal fields sensed by the muon ensemble below $T_{2}^{*}$ ${\simeq}$ 25 K. The increase in $\Gamma_\text{ZF}$ below $T_{2}^{*}$, measured at 1.6 K, is estimated to be ${\simeq}$ 0.045 $\upmu \rm s^{-1}$, which can be interpreted as a characteristic field strength $\Gamma_\text{ZF}$/${\gamma_{\mu}}$ ${\simeq}$ 0.45~G. 


%
\begin{figure*}
    \centering
    \includegraphics[width=1\linewidth]{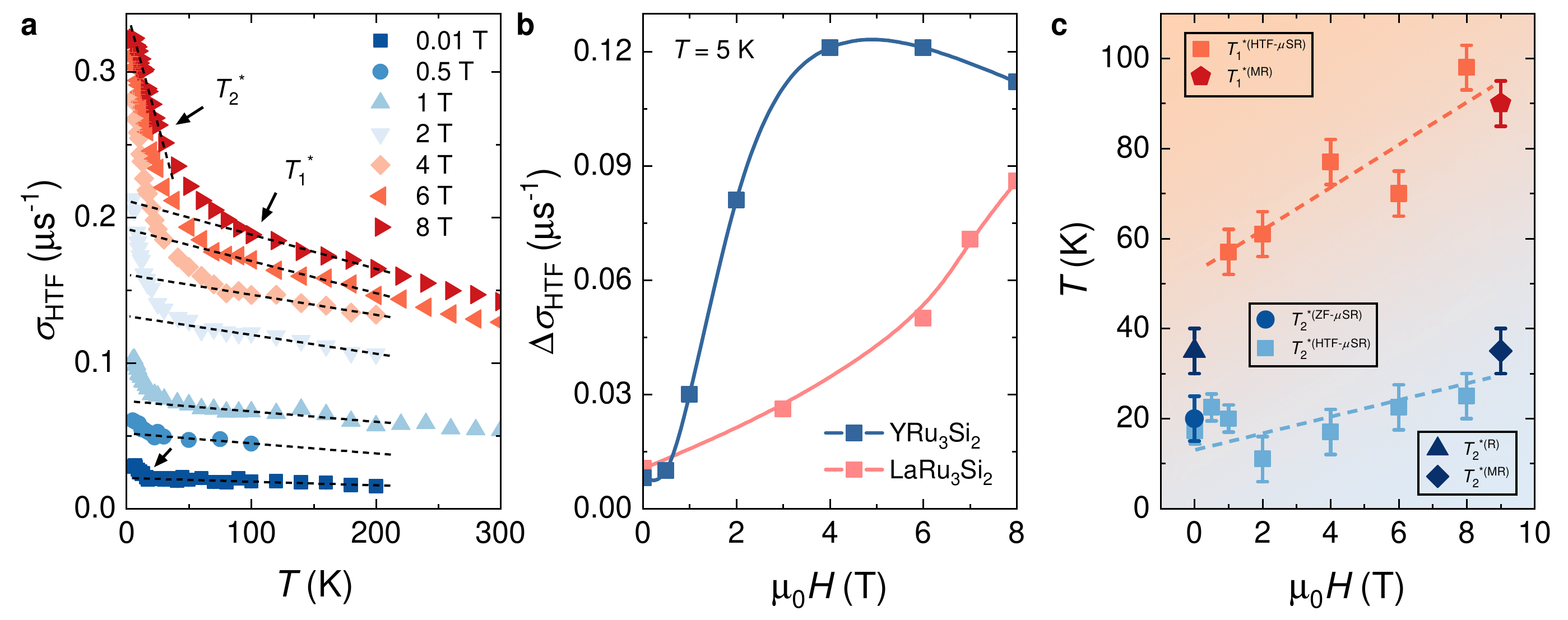}
    \caption{\textbf{Summary of the high-field $\mu$SR study of YRu$_{3}$Si$_{2}$ in the normal state.} \textbf{a}, Temperature dependence of the high transverse-field (HTF) muon spin relaxation rates measured for selected external fields (dashed black lines represent the linear fit of the high-temperature data). Arrows mark the characteristic temperatures \(T_{1}^*\) and \(T_{2}^*\). \textbf{b}, The low-temperature increase in the HTF $\mu$SR rate as a function of applied magnetic field, measured at 5 K. The data reported for LaRu$_{3}$Si$_{2}$\cite{mielke2024chargeordersdistinctmagnetic} analog are shown for comparison, as well. \textbf{c}, Characteristic temperatures \(T_{1}^*\) and \(T_{2}^*\) as determined from the HTF $\mu$SR, ZF $\mu$SR and magnetotransport measurements. Dashed lines are guides to the eye.}
    \label{fig:5}
\end{figure*}

\begin{figure*}
    \centering
    \includegraphics[width=1\linewidth]{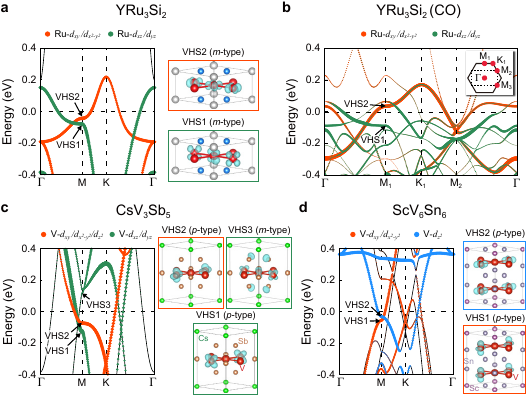}
    \caption{\textbf{VHS physics in YRu$_3$Si$_2$.} \textbf{a,c,d}, VHS points near the Fermi level and their characters for \textbf{a} YRu$_3$Si$_2$, \textbf{c} CsV$_3$Sb$_5$, and \textbf{d} ScV$_6$Sn$_6$. The band structure of the parent phase of each compound is shown, projected onto the $d$ orbitals of the V or Ru atom, with the radius of the open circles proportional to the projected weight. The charge density of VHS points is displayed where $m$-type and $p$-type refer to sublattice-mixed and sublattice-pure types, respectively. \textbf{b}, Unfolded band structure of the charge-ordered phase in YRu$_3$Si$_2$. In the inset, the solid line indicates the BZ of the parent $P6/mmm$ structure, while the dashed line indicates the BZ of the charge-ordered $Pmma$ structure. }
    \label{fig:VHS}
\end{figure*}

\begin{table*}
    \caption{\textbf{Summary of the charge order and time-reversal symmetry (TRS) breaking temperatures in kagome materials.} Comparison of the characteristic temperatures of charge order formation (\(T_\text{co}\)) and the temperatures below which the TRS is broken, either at zero ($T_\text{ZF}^{*}$) or high magnetic field ($T_\text{HF}^{*}$). The absolute increase in the zero-field muon spin relaxation rate below $T_\text{ZF}^{*}$, denoted as ${\Delta}$${\Gamma}_{\rm ZF}$, is also shown for various systems.}
    \centering
    \begin{tabular}{ccccccc}
    \hline
         Compound&  KV$_{3}$Sb$_{5}$\cite{mielke2022time}& RbV$_{3}$Sb$_{5}$\cite{guguchia2023tunable}& CsV$_{3}$Sb$_{5}$\cite{khasanov2022time}& ScV$_{6}$Sn$_{6}$\cite{arachchige2022charge,Guguchia2023hidden}&  LaRu$_{3}$Si$_{2}$\cite{mielke2024chargeordersdistinctmagnetic}&  YRu$_{3}$Si$_{2}$$^{\text{this study}}$\\
    \hline
         \(T_\text{co}\) (K)&  80& 110& 95&  80&  400&  800\\
         $T_\text{ZF}^{*}$ (K)&  80& 110, 50& 95, 30&  80&  35&  25\\
         $T_\text{HF}^{*}$ (K)&  80& 110& 95&  80&  80&  90\\
         ${\Delta}$${\Gamma}_{\rm ZF}$ (${\mu}s$$^{-1}$)&  0.025& 0.05& 0.04& 0.04& 0.035& 0.045\\
    \hline
    \end{tabular}
    \label{tab:SC-gap}
\end{table*}

To corroborate the zero-field ${\mu}$SR results presented above, a comprehensive set of high-field ${\mu}$SR experiments were conducted \cite{Sedlak2009}. The temperature dependencies of the muon spin relaxation rate $\sigma
_{\rm HTF}$, measured under various magnetic fields, are shown in Fig. \ref{fig:5}a. At the lowest applied field of 0.01 T, the relaxation rate remains nearly constant down to $T_{2}^{*}$, below which it exhibits a clear increase—consistent with the zero-field ${\mu}$SR results. The overall increase in 
$\sigma_{\rm HTF}$ below $T_{2}^{*}$ is smaller than that observed in zero field, as expected. As the applied field is increased, a stronger enhancement of the relaxation rate below $T_{2}^{*}$ is observed. At 4 T, the rate is nearly ten times larger than at 0.01 T, indicating a substantial field-induced enhancement of the magnetic response. Additionally, the increase in the rate below $T_{1}^{*}$ becomes more pronounced, and the temperature dependence above $T_{1}^{*}$ becomes increasingly evident. As shown in Figure \ref{fig:5}b, the absolute value of the increase saturates above 4 T, and the changes in the overall temperature dependence become less prominent. These measurements confirm the time-reversal symmetry (TRS) breaking nature of the phase below $T_{2}^{*}$, as well as the emergence of a field-induced magnetic state below $T_{1}^{*}$. Thus, muon spin rotation experiments reveal that the MR observed in YRu$_{3}$Si$_{2}$ originates from a hidden, weak magnetic state. A field-induced enhancement of the relaxation rate was previously observed in LaRu$_3$Si$_2$; however, the field dependence differs significantly between LaRu$_3$Si$_2$ and YRu$_3$Si$_2$, as shown in Fig. 5b. In YRu$_3$Si$_2$, the relaxation rate increases more rapidly at low fields and saturates above 4 T, whereas in LaRu$_3$Si$_2$, the rate continues to rise steadily up to 8 T. The characteristic temperatures $T_{1}^{*}$ and $T_{2}^{*}$, determined from high-field ${\mu}$SR, zero-field ${\mu}$SR, and magnetotransport measurements, are summarized in Fig. 5c, showing a smooth increase with applied field.
Furthermore, the continuous increase in the relaxation rate across a wide temperature range above $T_{1}^{*}$ suggests the presence of weak magnetic correlations at elevated temperatures, potentially linked to the high-temperature charge order. To verify this, future measurements extending up to the charge-ordering temperature $T_{\rm co}$ = 800 K will be necessary.

To summarize, our findings establish YRu$_3$Si$_2$ as a distinctive kagome-lattice superconductor exhibiting several remarkable properties: (1) High-temperature charge order below $T_{\rm co}$ ${\simeq}$ 800 K: YRu$_3$Si$_2$ demonstrates a record-high onset temperature for charge order, characterized by a propagation vector of (1/2, 0, 0).
This charge order wave vector differs from the (1/4) and (1/6) modulations observed in LaRu$_3$Si$_2$, but closely resembles the one reported in AV$_3$Sb$_5$ systems \cite{wilson2024v3sb5}. (2) Time-reversal symmetry-breaking Phase below $T_{2}^{*}$ ${\simeq}$ 25 K: Between the onset of charge order and the superconducting transition, an electronically driven phase that break TRS emerges, indicating complex underlying electronic interactions. TRS breaking in the normal state has been reported in several charge-ordered kagome systems, including AV$_3$Sb$_5$, ScV$_6$Sn$_6$, and LaRu$_3$Si$_2$. Furthermore, the magnitude of the TRS-breaking signal is quite comparable across different kagome systems (see Table 1). This indicates that this phenomenon is a ubiquitous feature of the kagome lattice. However, a notable distinction emerges: in ScV$_6$Sn$_6$, the onset of TRS breaking coincides closely with the onset of charge order. In AV$_3$Sb$_5$, the onset of TRS breaking coincides with the emergence of charge order, with a stronger TRS-breaking signal developing at lower temperatures (see Table 1). In contrast, a notable difference arises in LaRu$_3$Si$_2$ and YRu$_3$Si$_2$, where TRS breaking sets in at temperatures well below the primary charge-ordering transition (see Table 1). (3) Field-induced magnetic response at least below $T_{1}^{*}$ ${\simeq}$ 90 K. (4) Band structure calculations identify flat band and two VHSs near the Fermi level (see detailed characterization of the VHSs below), suggesting a strong interplay between electronic correlations and emergent orders.
(5) Superconductivity with an unconventional gap structure: The superconducting state in YRu$_3$Si$_2$ features a gap structure consistent with either two isotropic full gaps or an anisotropic nodeless gap, suggesting unconventional pairing mechanisms. This makes YRu$_3$Si$_2$ an ideal platform for exploring the interplay between multiple symmetry-breaking phases and their connection to superconductivity, and it strongly motivates further investigations using techniques such as STM, ARPES, and beyond.
 
Finally, we discuss the possible origin of the observed time-reversal symmetry breaking phase within the charge-ordered phase. Two VHSs are identified near the Fermi level in the pristine band structure, one of which lies within the flat band (Fig. \ref{fig:VHS}a). These VHSs exhibit orbital characters of the out-of-plane $d_{xz}$ and $d_{yz}$ orbitals for VHS1 and the in-plane $d_{xy}$ and $d_{x^2-y^2}$ orbitals for VHS2. This is reminiscent of the VHSs with both in-plane and out-of-plane orbital characters in CsV$_3$Sb$_5$ (Fig. \ref{fig:VHS}c) and ScV$_6$Sn$_6$ (Fig. \ref{fig:VHS}d), where time-reversal symmetry breaking orders have been reported\cite{mielke2022time,Guguchia2023hidden}. Closer inspection shows that both VHSs in YRu$_3$Si$_2$ belong to sublattice-mixed type, consisting of a mixed contribution from two kagome sublattices (Fig. \ref{fig:VHS}a). This is distinct from both the sublattice-mixed and sublattice-pure VHSs in CsV$_3$Sb$_5$ (Fig. \ref{fig:VHS}c), as well as from the sublattice-pure VHSs in ScV$_6$Sn$_6$ (Fig. \ref{fig:VHS}d). Interestingly, we reveal that the VHSs characteristics in the pristine phase are preserved in the charge-ordered state, as shown in the unfolded band structure in Fig. \ref{fig:VHS}b. Upon the charge order formation, the bandwidths of the two VHS bands are renormalized along the $\Gamma-\text{M}_1-\text{K}_1$ direction, positioning VHS1 (VHS2) below (above) the Fermi level at the M$_1$ point, while retaining their orbital characters. This is particularly remarkable, as recent theoretical studies suggest that the presence of two VHSs near the Fermi level is a key prerequisite for loop current order in kagome metals\cite{Christensen2022,Li2024intertwined}. We thereby attribute the observed time-reversal symmetry breaking phase within the charge-ordered phase to the persistent multiple VHSs, even under the charge order formation. YRu$_3$Si$_2$ serves as a distinct example, with different VHS characters compared to CsV$_3$Sb$_5$ and ScV$_6$Sn$_6$, which may contribute to uncovering the exact mechanism of loop current order in the kagome metal family. 

We also note that the lowered symmetry of the charge-ordered state inherently gives rise to nematicity in VHS fermiology of YRu$_3$Si$_2$ (Fig. \ref{fig:VHS}b). Compared to the two resilient VHSs at the M$_1$ point with some bandwidth renormalization, the VHSs at the M$_2$ point are washed out due to charge-order driven interactions between multiple VHSs originating from the M$_2$ and M$_3$ points in the pristine BZ. 
This VHS fermiology is clearly distinct from the VHSs present at all three M points in the pristine phase of CsV$_3$Sb$_5$, but it is similar to the nematicity-driven lifting of VHSs, leaving VHSs only at a single M point in the charge-ordered  phase of ScV$_6$Sn$_6$ below the nematic temperature\cite{Jiang2024VanHove}. 
The role of symmetry breaking and its impact on exotic quantum phases in kagome metals is of crucial importance, particularly in the context of VHS-driven instabilities\cite{Kim2023_monolayer,Guo2024correlated}.
Thus, YRu$_3$Si$_2$ presents itself as an intriguing kagome material for studying the interplay between VHSs and nematicity in the formation of time-reversal symmetry breaking loop current order and superconductivity.

\section*{Methods}

\textbf{Sample preparation:}
Polycrystalline samples of YRu$_{3}$Si$_{2}$ were prepared by arc-melting from mixtures of yttrium chunks (99.9 ${\%}$, Thermo Scientific), ruthenium powder (99.99 ${\%}$, Leico), and silicon pieces (99.95 ${\%}$, Sigma Aldrich) under an argon atmosphere. Stoichiometric amounts of the elements were used with a 30 ${\%}$ molar excess of ruthenium to avoid the formation of the YRu$_{3}$Si$_{2}$ phase. The ruthenium powder was pressed into pellets to avoid sputtering, and the melting process was started with the metals (yttrium and ruthenium) so that the melt could absorb silicon. A piece of zirconium was used as a getter to remove residual oxygen. During the arc-melting process, the samples were melted and flipped several times for better homogenization. The pellets were not shiny on the surface after the synthesis, this was due to the formation of a thin oxide layer, which was mechanically removed.\\

\textbf{First-principles calculations:}
We perform density functional theory (DFT) calculations using the Vienna ab initio simulation package {\sc vasp} \cite{VASP} implementing the projector-augmented wave method \cite{PAW}. We use PAW pseudopotentials with valence configurations: $4s^2 4p^6 4d^2 5s^1$ for Y atoms, $4s^2 4p^6 4d^7 5s^1$ for Ru atoms, and $3s^2 3p^2$ for Si atoms. We approximate the exchange correlation functional with the generalized-gradient approximation PBEsol \cite{PBEsol}. We use a kinetic energy cutoff for the plane wave basis of 400\,eV and a Gaussian smearing of 0.02\,eV. We use $\Gamma$-centered \textbf{k}-point grids with a \textbf{k}-spacing of $0.1$\,$\text{\AA}^{-1}$. All the structures are optimized until the forces are below 0.001\,eV/\AA. The unfolded band structure of the charge-ordered phase is obtained using the method proposed by Popescu and Zunger \cite{popescu2012extracting} as implemented in
the VaspBandUnfolding code \cite{vaspbandunfolding}. We perform harmonic phonon calculations using the finite displacement method in conjunction with nondiagonal supercells \cite{nondiagonal_supercells}. The dynamical matrices are calculated on uniform \textbf{q} grids of size $6\times6\times6$ for the high-temperature $P6/mmm$ structure, and of size $2\times2\times2$ for the $Cccm$ and $Pmma$ structures. We find that both the $Cccm$ and $Pmma$ structures are dynamically stable at the harmonic level (Extended data, Figure 7).\\


\textbf{Single crystal X-ray diffraction measurements:}

Single crystals of YRu$_3$Si$_2$ of up to 50 ${\upmu}$m were selected from a crushed arc-melted button, washed consequently in dilute HCl, H$_{2}$O and ethanol and mounted on MiTeGen loops or a quartz capillary. Diffraction measurements were carried out using the laboratory STOE STADIVARI single crystal diffractometer equipped with the micro-focused Mo $K_{\alpha}$ X-ray source and Dectris EIGER 1M 2R detector. Measurements were carried out in the temperature range between 80 and 400 K using Oxford CryoStream 800 Series and between RT and 900 K using STOE HeatStream (Ar flow of 0.8 L/min). Data collection and reduction as well as reconstruction of reciprocal space were performed using X-Area software package [X-area package (STOE and Cie GmbH, Darmstadt, Germany, 2022)]. Structure solution and refinement were performed using Jana2020 software \cite{PetříčekPalatinusPlášilDušek+2023+271+282}. Similar to LaRu$_3$Si$_2$, hexagonal pseudo-symmetry of the diffraction patterns (Fig. 1) is due to ortho-hexagonal twinning, which was accounted for during refinement of the lt-hex and charge ordered phases.\\

\textbf{Magnetotransport measurements:}
Magnetotransport measurements were carried out in a standard four-probe method using the Physical Property Measurement System (PPMS, Quantum Design). The diagonal arrangement of voltage contacts was used and the magnetoresistance and Hall resistance were obtained by symmetrization and anti-symmetrization of the measured data, respectively.\\

\textbf{Muon Spin Rotation Experiments:}
Zero-field (ZF) and transverse-field (TF) $\mu$SR experiments were performed on the GPS instrument and high-field HAL-9500 instrument, equipped with BlueFors vacuum-loaded cryogen-free dilution refrigerator (DR), at the Swiss Muon Source (S$\mu$S) at the Paul Scherrer Institut, in Villigen, Switzerland.

In a ${\mu}$SR experiment nearly 100 ${\%}$ spin-polarized muons $\mu$$^{+}$ are implanted into the sample one at a time. The positively charged $\mu$$^{+}$ thermalize at interstitial lattice sites, where they act as magnetic microprobes. In a magnetic material, the muon spin precesses in the local field $B_{\rm \mu}$ at with the Larmor frequency ${\nu}_{\rm \mu}$ = $\gamma_{\rm \mu}$/(2${\pi})$$B_{\rm \mu}$ (muon gyromagnetic ratio $\gamma_{\rm \mu}$/(2${\pi}$) = 135.5 MHz T$^{-1}$). Using the $\mu$SR technique, important length scales of superconductors can be measured, namely the magnetic penetration depth $\lambda$ and the coherence length $\xi$. If a type-II superconductor is cooled below $T_{\rm c}$ in an applied magnetic field ranged between the lower ($H_{c1}$) and the upper ($H_{c2}$) critical fields, a vortex lattice is formed which in general is incommensurate with the crystal lattice with vortex cores separated by much larger distances than those of the unit cell. Because the implanted muons stop at given crystallographic sites, they will randomly probe the field distribution of the vortex lattice. Such measurements need to be performed in a field applied perpendicular to the initial muon spin polarization (the so-called TF configuration).

Zero-field and high-transverse-field experiments were performed to probe normal state properties on the HAL instrument in a field range of 0.01 to 8 T. The sample in the form of compressed pellet (diameter 8 mm) was placed on the silver sample holder and mounted in the cryostat.\\

\textbf{Analysis of TF-${\mu}$SR data}:

In order to model the asymmetric field distribution ($P (B)$) in the SC state, the TF-${\mu}$SR time spectra measured below $T_{\rm c}$ are analyzed using the following two-component functional form:

\begin{equation}
	\begin{aligned}
		A_{\rm TF} (t) = \sum_{i=0}^{2} A_{s,i}\exp\Big[-\frac{\sigma_{i}^2t^2}{2}\Big]\cos(\gamma_{\mu}B_{{\rm int},s,i}t+\varphi)  
		\label{eqS1}
	\end{aligned}
\end{equation}

Here $A_{s,i}$, $\sigma_{i}$ and $B_{{\rm int},s,i}$ is the initial asymmetry, relaxation rate, and local internal magnetic field of the $i$-th component. ${\varphi}$ is the initial phase of the muon-spin ensemble. $\gamma_{\mu}/(2{\pi})\simeq 135.5$~MHz/T is the muon gyromagnetic ratio. The first and second moments of the local magnetic field distribution are given by~\cite{Khasanov104504}

\begin{equation}
	\begin{aligned}
		\left\langle {B}\right\rangle = \sum_{i=0}^{2} \frac {A_{s,i}B_{{\rm int},s,i}}{A_{s,1}+A_{s,2}}
		\label{eqS2}
	\end{aligned}
\end{equation}
and 
\begin{equation}
	\begin{aligned}
		\left\langle {\Delta B}\right\rangle ^2 = \frac {\sigma ^2}{\gamma_{\mu}^2} =  \sum_{i=0}^2  \frac {A_{s,i}}{A_{s,1}+A_{s,2}} \Big[\sigma_i^2/\gamma_{\mu}^2 + \left(B_{{\rm int},s,i} - \langle B \rangle\right)^2\Big].
	\end{aligned}
\end{equation}

Above $T_{\rm c}$, in the normal state, the symmetric field distribution could be nicely modeled by only one component. The obtained relaxation rate and internal magnetic field are denoted by $\sigma_{\rm ns}$ and $B_{{\rm int},s,{\rm ns}}$. $\sigma_{\rm ns}$ is found to be small and temperature independent (dominated by nuclear magnetic moments) above $T_{\rm c}$ and assumed to be constant in the whole temperature range. Below $T_{\rm c}$, in the SC state, the relaxation rate and internal magnetic field are indicated by $\sigma_{\rm SC}$ and $B_{{\rm int},s,{\rm sc}}$. $\sigma_{\rm SC}$ is extracted by using $\sigma_\mathrm{SC} = \sqrt{\sigma^{2} - \sigma^{2}_\mathrm{ns}}$. $B_{{\rm int},s,{\rm sc}}$ is evaluated from $\left\langle {B}\right\rangle$ using Eq.~\eqref{eqS2}.\\

\textbf{Analysis of ${\lambda}(T)$}:

${\lambda}$($T$) was calculated within the local (London) approximation (${\lambda}$ ${\gg}$ ${\xi}$) by the following expression \cite{Suter69, Tinkham2004}:
\begin{equation}
	\frac{\sigma_{\rm SC}(T,\Delta_{0,i})}{\sigma_{\rm SC}(0,\Delta_{0,i})}=
	1+\frac{1}{\pi}\int_{0}^{2\pi}\int_{\Delta(_{T,\varphi})}^{\infty}(\frac{\partial f}{\partial E})\frac{EdEd\varphi}{\sqrt{E^2-\Delta_i(T,\varphi)^2}},
\end{equation}
where $f=[1+\exp(E/k_{\rm B}T)]^{-1}$ is the Fermi function, ${\varphi}$ is the angle along the Fermi surface, and ${\Delta}_{i}(T,{\varphi})={\Delta}_{0,i}{\Gamma}(T/T_{\rm c})g({\varphi}$)
(${\Delta}_{0,i}$ is the maximum gap value at $T=0$). The temperature dependence of the gap is approximated by the expression ${\Gamma}(T/T_{\rm c})=\tanh{\{}1.82[1.018(T_{\rm c}/T-1)]^{0.51}{\}}$,\cite{Carrington205} while $g({\varphi}$) describes the angular dependence of the gap and is replaced by 1 for both a single isotropic gap and and two isotropic full gaps, and ${\mid}\cos(2{\varphi}){\mid}$ for a nodal $d$ wave gap~\cite{Guguchia8863}.\\


\subsection*{Data availability}
All related data are available from the authors on request. 

\subsection*{Acknowledgments}
The ${\mu}$SR experiments were carried out at the Swiss Muon Source (S${\mu}$S) Paul Scherrer Insitute, Villigen, Switzerland. Z.G. acknowledges support from the Swiss National Science Foundation (SNSF) through SNSF Starting Grant (No. TMSGI2${\_}$211750). Z.G. acknowledges the useful discussions with Robert Scheuermann. S.-W.K. acknowledges support from a Leverhulme Trust Early Career Fellowship (ECF-2024-052). K.W., B.M., and S.-W.K. acknowledge support from a UKRI Future Leaders Fellowship [MR/V023926/1]. I.P. acknowledges support from Paul Scherrer Institute research grant No. 2021 01346. The computational resources were provided by the Cambridge Tier-2 system operated by the University of Cambridge Research Computing Service and funded by EPSRC [EP/P020259/1], and by the UK National Supercomputing Service ARCHER2, for which access was obtained via the UKCP consortium and funded by EPSRC [EP/X035891/1]. Z.W.  is supported by the U.S. Department of Energy, Basic Energy Sciences Grant DE-FG02-99ER45747.

\section{Author contributions}~
Z.G. conceived, designed, and supervised the project. Density Functional Theory Calculations: S.-W.K., K.W., and B.M.. Sample growth: A.L., M.S. and F.v.R.. Magnetotransport experiments: V.S. and Z.G..
Laboratory X-ray diffraction experiments: I.P., D.G., and Z.G.. X-ray diffraction experiments at DESY: I.B., O.G., P.K., L.M., J.O., M.V.Z. and J.C.. Muon Spin rotation experiments and corresponding discussions: P.K., J.N.G., V.S., O.G., S.S.I., A.D., H.L., R.K., J.-X. Yin, Z.W. and Z.G.. Data analysis, figure development and writing of the paper: P.K., S.-W.K., and Z.G. with contributions from all authors. All authors discussed the results, interpretation, and conclusion.

\bibliography{References}{}

\begin{figure*}
    \centering
    \includegraphics[width=1.0\linewidth]{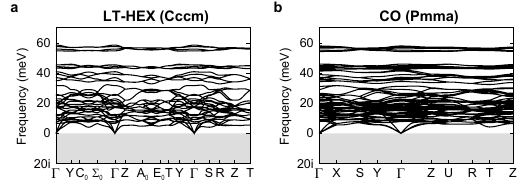}
    \vspace{-0.5cm}
    \caption{\textbf{Phonon dispersions of LT-HEX and CO states in YRu$_{3}$Si$_{2}$.} \textbf{a-b}, Phonon dispersions of \textbf{a} LT-HEX ($Cccm$) and \textbf{b} CO ($Pmma$) phases in YRu$_{3}$Si$_{2}$. Both structures are dynamically stable. 
    }
    \label{Figure1}
\end{figure*}

\end{document}